\documentclass[fleqn,usenatbib]{mnras}

\usepackage{newtxtext,newtxmath}

\usepackage[T1]{fontenc}
\usepackage{ae,aecompl}

\usepackage{graphicx}	
\usepackage{amsmath}	
\usepackage{amssymb}	
\usepackage{verbatim}
\usepackage[flushleft]{threeparttable}
\usepackage[dvipsnames]{xcolor}
\usepackage{soul}

\newcommand{\qso}{Q\,0528$-$250}
\newcommand{\zem}{\ensuremath{z_{\rm em}}}
\newcommand{\zabs}{\ensuremath{z_{\rm abs}}}
\newcommand{\HI}{H\,{\sc i}}
\newcommand{\CI}{C\,{\sc i}}
\newcommand{\SiII}{Si\,{\sc ii}}
\newcommand{\kmps}{\rm km\,s$^{-1}$}

\definecolor{green}{rgb}{0,0.4,0}

\newcommand{\ioffe}{Ioffe Institute, {Polyteknicheskaya 26}, 194021 Saint-Petersburg, Russia}
\newcommand{\eso}{European Southern Observatory, Alonso de C\'ordova 3107, Vitacura, Casilla 19001, Santiago, Chile}
\newcommand{\iap}{Institut d'Astrophysique de Paris, CNRS-SU, UMR\,7095, 98bis boulevard Arago, 75014 Paris, France}
\newcommand{\iucaa}{Inter-University Centre for Astronomy and Astrophysics, Pune University Campus, Ganeshkhind, Pune 411007, India}

\title[Nature of the gas at $z_{\rm abs}>z_{\rm em}$ towards Q\,0528-250]{Nature of the DLA towards Q\,0528$-$250\thanks{Based on data collected at the European Southern
Observatory under ESO programmes 66.A-0594, 68.A-0600, 68.A-0106, and 082.A-0087.}:\\
High pressure and strong UV field revealed by excitation of
\ion{C}{i}, H$_2$ and \ion{Si}{ii}
}

\author[S.~A.~Balashev et al.]{S.~A.~Balashev,$^{1}$\thanks{E-mail: s.balashev@gmail.com}
C.~Ledoux,$^{2}$
P.~Noterdaeme,$^{3}$
R.~Srianand,$^{4}$ 
P.~Petitjean,$^{3}$ \newauthor
and N.~Gupta$^{4}$
\\
$^{1}$ \ioffe \\
$^{2}$ \eso \\
$^{3}$ \iap \\
$^{4}$ \iucaa \\
}

\date{Accepted 2020 July 15. Received 2020 June 28; in original form 2020 March 6}

\pubyear{}

\begin{document}
\label{firstpage}
\pagerange{\pageref{firstpage}--\pageref{lastpage}}
\maketitle

\begin{abstract}
We present the detection of excited fine-structure energy levels of singly-ionized silicon
and neutral carbon associated with the proximate damped Lyman-$\alpha$ system at
$z_{\rm abs}=2.811$ towards \qso. This absorber has an apparent relative velocity that
is inconsistent with the Hubble flow indicating motion along the line-of-sight towards
the quasar, i.e., $\zabs>\zem$. We measure the metallicity of the system to
be ${\rm [Zn/H]}=-0.68\pm 0.02$. Using the relative populations of the fine-structure levels
of \ion{Si}{ii} and \ion{C}{i}, as well as the populations of H$_2$ rotational levels,
we constrain the physical conditions of the gas. We derive { hydrogen} number
densities of $n_{\rm H}=190^{+70}_{-50}$\,cm$^{-3}$ and $260^{+30}_{-20}$\,cm$^{-3}$ in
two velocity components where both \ion{C}{i} and H$_2$ are detected. Taking into
account the kinetic temperature in each component, $\sim 150$\,K, we infer high values of
thermal pressure in the cold neutral medium probed by the observations. The strengths of
the UV field in Draine's unit are $I_{\rm UV} = 10^{+5}_{-3}$ and $14^{+3}_{-3}$ in each
of these two components, respectively. Such enhanced UV fluxes and thermal pressure compared
to intervening DLAs are likely due to the proximity of the quasar. The typical size of
the absorber is $\sim 10^4$~a.u. Assuming the UV flux is dominated by the quasar, we
constrain the distance between the quasar and the absorber to be $\sim 150-200$\,kpc.
This favours a scenario where the absorption occurs in a companion galaxy located in
the group where the quasar-host galaxy resides. This is in line with studies in emission
that revealed the presence of several galaxies around the quasar.
\end{abstract}




\section{Introduction}

Quasars are active phases of super-massive black holes that reside in the core of galaxies and play a
major role in the formation and evolution of galaxies through so-called feedback processes. Quasars are
fed with gas principally released through galaxy interactions and/or mergers. The rate of such processes
is significantly enhanced in denser environments and therefore quasar hosts are believed to be
predominantly located in massive galaxy groups. The gas in these groups can be studied in absorption in the
spectra of quasars that may either be group members or located in the background. Of particular interest are
the proximate DLAs (i.e., PDLAs), i.e., damped Lyman-$\alpha$ systems (DLAs;
with $\log N($\ion{H}{i}$)\gtrsim 20$) associated with the quasar host galaxy and/or its environment. These
absorbers allow to explore the physical conditions of high column-density gas under the influence of
quasar feedback.

Based on the CORALS survey, \citet{Ellison2002} reported a factor of $\sim 4$ excess of PDLAs compared to
intervening DLAs. \citet{Prochaska2008} found a factor of $\sim2$ excess at redshifts $z\sim3$ from the
Sloan Digital Sky Survey (SDSS) data release-5, but no statistically significant excess at $z<2.5$ or $z>3.5$.
Additionally, \citet{Noterdaeme2019} found that H$_2$-selected PDLAs with $z>3$ show an excess by factors
of 4-5\footnote{The exact value depends on the velocity range used to define PDLAs, possibly reaching
an excess of an order of magnitude within 1000~\kmps\ from the quasar systemic redshift.} compared
to intervening DLAs selected in the same way. The discrepancy between these studies may be due to the
complex spectral-line profile in the Ly$\alpha$ region for a large fraction of PDLAs
\citep[e.g.,][]{Hennawi2009, Finley2013}, caused by strong residual Ly$\alpha$ emission not absorbed by
the DLA. Such PDLAs are called coronagraphic. PDLA searches based on the standard assumption of complete
Ly$\alpha$ absorption troughs may miss a large fraction of such systems.

In extreme cases, the extent of the neutral gas clouds producing the DLA is smaller than the size of the Broad Line Region (BLR) of the quasar. The leaking broad Ly$\alpha$ emission may then fill the DLA trough, resulting in the formation of a "ghostly" DLA in the quasar spectrum \citep{Jiang2016, Fathivavsari2017}. Usually, coronagraphic DLAs exhibit relatively high excitation
of fine-structure energy levels of Si\,{\sc ii}, C\,{\sc i}, and/or O\,{\sc i}. Based on the observation of a
direct relation between the strength of leaking Ly$\alpha$ emission and fine-structure excitation of metal
species, \citet{Fathivavsari2018} suggested that systems with strong Ly$\alpha$ emission (and therefore with
probably less covering of the Ly$\alpha$ emitting region) could be denser and located closer to the AGN.
Such DLAs could be subject to mechanical compression of the gas and/or enhanced incident UV fluxes, both of
which can contribute to the excitation of the above-mentioned fine-structure levels.

{ Though these distinct classes of PDLAs are now routinely identified in SDSS
spectra, the spectral resolution of the SDSS ($R\sim 2000$) is insufficient to study them
in detail. Medium- to high-resolution observations are required to exploit the
information encoded in the fine-structure levels of metal species and/or H$_2$
rotational levels, which can then be used to infer the excitation and physical conditions
of the gas in these harsh environments. Unfortunately, follow-up observations are
difficult as most of the corresponding quasars are faint and, to date, only a
few high-resolution spectra of PDLAs towards relatively bright quasars are available. Among
those, the DLA system at $z_{\rm abs}=2.811$ towards \qso\ stands out as a unique
absorber for having a redshift higher than that of the quasar by several thousands
of km~s$^{-1}$, suggesting proximity to the quasar, together with the detection of
prominent H$_2$ lines.}

{ In this paper, we focus on the method to constrain the fine-structure excitation of the
\ion{Si}{ii} and \ion{C}{i} ground states and analyse the physical conditions of the
absorbing gas. In Sect.~\ref{sect:history}, we summarize previous studies of
this DLA. Using high-quality VLT spectra retrieved from the ESO archive
\citep[see][]{Klimenko2015}, we confirm the presence of \ion{C}{i} and \ion{C}{i*}
and present new detection of \ion{Si}{ii*} and \ion{C}{i**} in this system
(Sect.~\ref{sect:analysis}). Coupled with a study of H$_2$ rotational excitation, this
allows us to estimate the hydrogen number density and UV flux in the cold neutral
medium associated with the H$_2$ components and to constrain the physical distance between
the DLA and the quasar (Sect.~\ref{sect:PhysCond}). We discuss the implications of
this result in Sect.~\ref{sect:discussion} and conclude in Sect.~\ref{sect:conclusion}.}

\section{Previous studies of \qso}

\label{sect:history}
The DLA system at $z_{\rm abs}=2.811$ towards \qso\ was identified by \citet{Jauncey1978} from follow-up low-resolution optical  spectroscopy ($\Delta\lambda$=10\,\AA) of radio sources from the Parkes 2.7~GHz sample. \citeauthor{Jauncey1978} did not detect any emission line in this spectrum, but subsequent observations by \citet{Smith1979} revealed \ion{Si}{iv} and \ion{C}{iv} emission lines at $z=2.765\pm0.01$, i.e., significantly lower than the DLA redshift, which indicated that the absorber is located in the vicinity of the quasar. A higher resolution spectrum ($\Delta\lambda$=2\,\AA) of the quasar was obtained by \citet{Morton1980} and later on used by \citet{Levshakov1985} who reported the first ever detection of molecular-hydrogen absorption lines at high redshift. This was confirmed by \citet{Foltz1988} from $\Delta\lambda$=1\,\AA\ spectroscopy and they derived
$\log N(\rm H_2)\sim18$ (column densities will be expressed in cm$^{-2}$ throughout).
The first high-resolution Echelle spectrum (R$\sim$36\,000) of \qso\ was obtained by \citet{Cowie1995} using HIRES at Keck, and then and a number of other studies  aiming to constrain the possible time variation of the proton-to-electron mass ratio from analysis of H$_2$ absorption lines \citep{Foltz1988,Varshalovich1995,Potekhin1998,Ubachs2004,King2008,King2011}.

\citet{Ge1997} used a MMT spectrum ($\Delta\lambda$=2\,\AA) of \qso\ to study the physical conditions in the medium through modeling of the abundances of ionized species. They argued that the number density is about 20\,cm$^{-3}$, the physical size of the components is $\sim40$\,pc, and the distance of the absorber to the quasar is larger than 1~Mpc. However, most of the absorption lines measured in this spectrum are not resolved and are saturated, meaning that the column densities and hence the physical conditions of the gas may have been incorrectly derived.
A high value of the number density in the H$_2$-bearing gas, $n\sim 1000$\,cm$^{-3}$, was reported by \citet{Srianand1998} from the analysis of a spectrum obtained using CASPEC at the 3.6~m telescope of La~Silla observatory. To derive this number density, the latter authors used the observed relative populations of H$_2$ rotational levels up to $J=4$. However, self-shielding was not
taken into account and the
H$_2$ column densities reported in this work are two orders of magnitude smaller than those derived at higher spectral resolution. The reported H$_2$ column densities indeed vary significantly between different analysis ($\log N(\rm H_2)\sim 16.5$ in \citealt{Levshakov1985,Ge1997,Srianand1998,King2011} and $\log N(\rm H_2)\sim 18$ in \citealt{Foltz1988,Ledoux2003,Klimenko2015}).

The high signal-to-noise, high-resolution spectrum obtained in 2002-2003 by \citet{Ledoux2003} using the UVES spectrograph mounted at the VLT made it possible to resolve two components in the H$_2$ absorption-line profiles, which favoured
a high H$_2$ column-density value \citep{Srianand2005}. Using this spectrum, these authors also detected \CI\ lines from the first-two fine-structure energy levels of the atom ground state, associated with one of the H$_2$ components. This allowed them to revise the number density and thermal pressure of the H$_2$/\CI-bearing gas to be $n_{\rm H}\approx 25-270$\,cm$^{-3}$ and $P\approx 3250-17\,000$\,cm$^{-3}$K, respectively.
\citet{Cirkovic2006} used the same UVES spectrum to derive an upper limit on the column density of HD molecules, which resulted in loose constraints on the cosmic-ray ionizing flux.
The detection of HD absorption in this system was reported by \citet{King2011}, who measured $\log N(\rm HD)=13.27\pm0.07$ from new UVES observations carried out in 2009. The presence of \CI, H$_2$, and HD, suggests that the line-of-sight intercepts cold gas. Many observational campaigns aimed at detecting 21\,cm absorption towards \qso\ \citep[e.g.,][]{Carilli1996,Curran2010,Kanekar2014}, but they all led to lower limits on the ratio of the spin temperature to the covering fraction, $T_s/f>700$\,K. Using these limits, \citet{Srianand2012} estimated the column density of \HI\ associated with H$_2$, $\log N($\HI$)\sim 20$, and constrained the size of the clouds to be $<1.3$\,pc. Finally, \citet{Klimenko2015} used the combination of all available UVES observations to confirm the detection of HD and also noticed partial covering of the QSO-continuum emission region by the H$_2$ gas, with a covering fraction of $\sim$98\%.

In addition to absorption-line studies, the field of the $z_{\rm abs}=2.811$ DLA towards \qso\ was also searched
for in emission. This is indeed one of the first DLAs where Ly$\alpha$ emission was detected. Using narrow-band
imaging with EFOSC mounted at the ESO 3.6\,m telescope, \citet{Moller1993} detected three Ly$\alpha$-emitting sources
at $z_{\rm em}\simeq 2.8$ with impact parameters of $\sim$1, 11, and 21~arcsec (corresponding to 8, 88, 168 kpc of transverse distance), from the quasar. \citet{Warren1996}
confirmed these findings and measured the redshifts, widths, and fluxes, of Ly$\alpha$-emission lines from long-slit
spectroscopy.
Subsequent HST imaging and NTT/SUSI narrow-band imaging allowed to perform photometry and measure the sizes of the
Ly$\alpha$-emitting galaxies \citep{Moller1998}. The latter authors found that the source with the largest
impact parameter is in fact made of two components and that its Ly$\alpha$ emission is much more extended than
the continuum. This led them to argue that: (i) the Ly$\alpha$-emitting sources correspond to a group of galaxies in
the vicinity of the quasar; (ii) the Ly$\alpha$ emission of these sources is probably due to scattered emission from the quasar rather
than in-situ star-formation; (iii) the source with the smallest impact parameter is most likely the
DLA-galaxy counterpart. However, since radiative transfer of Ly$\alpha$ photons is complex, it is hard to constrain
the physical separation between these galaxies and the quasar from Ly$\alpha$ emission.

\begin{table}
\caption{Fit results for \ion{C}{i} ground-state and fine-structure lines originating from the first two
excited levels of singly-ionized carbon. Total column densities, either across all components for a given
fine-structure energy level, or summing up the contributions of all fine-structure levels in a given component,
are provided.}
\begin{center}
\begin{tabular}{cccc}
\hline
Component & $A$ & $B$ & $A+B$ \\
\hline
$z$ & $2.810973(^{+6}_{-6})$ & $2.8111231(^{+15}_{-12})$ &  \\
$\Delta v^\dagger$ [km~s$^{-1}$] & $-11.9^{+0.5}_{-0.5}$             & $-0.1^{+0.1}_{-0.1}$                 &  \\
$b$        [km~s$^{-1}$] & $3.8^{+1.1}_{-1.1}$ & $1.9^{+0.4}_{-0.4}$ &  \\
\hline
$\log N($\ion{C}{i}$)$              & $11.65^{+0.06}_{-0.06}$ & $12.06^{+0.02}_{-0.02}$ & $12.20^{+0.02}_{-0.02}$ \\
$\log N($\ion{C}{i}$^\star)$        & $11.83^{+0.07}_{-0.07}$ & $12.27^{+0.02}_{-0.03}$ & $12.40^{+0.02}_{-0.02}$ \\
$\log N($\ion{C}{i}$^{\star\star})$ & $<11.34$                & $11.95^{+0.03}_{-0.03}$ & $11.99^{+0.04}_{-0.04}$ \\
\hline
$\log N_{\rm tot}$                  & $12.07^{+0.07}_{-0.06}$ & $12.59^{+0.02}_{-0.01}$ & $12.71^{+0.02}_{-0.02}$ \\
\hline
\end{tabular}
\begin{tablenotes}
\item $^\dagger$ Velocity offset relative to $z=2.811124$.
\end{tablenotes}
\end{center}
\label{table:CI_results}
\end{table}

\section{Spectroscopic analysis}
\label{sect:analysis}

In the present study, we use the combination of all exposures obtained with the Ultraviolet and Visual Echelle
Spectrograph \citep[UVES;][]{Dekker2000} on the Very Large Telescope during two epochs of observations, in
2001-2002, and in 2008-2009. The details of the observations and the data reduction are described in
\citet{Klimenko2015}.

We fit neutral atomic-hydrogen absorption lines using a one-component model (since the Ly$\alpha$ line is
strongly damped) together with the unabsorbed quasar continuum modelled using six Chebyshev
polynomials \citep[for details on the technique, see, e.g.,][]{Balashev2019}. We find that the total
neutral-hydrogen column density is $\log N($\ion{H}{i}$)=21.37\pm 0.01$ centered on
$z_{\rm abs}=2.811379(20)$ in agreement with the value reported by \citet{Ledoux2006}
($\log N($\ion{H}{i}$)=21.35\pm 0.07$). The Ly$\alpha$ absorption line is redshifted
by $\sim 2000$~\kmps compared to the reconstructed profile of the Ly$\alpha$ emission line. However, it
is known that Ly$\alpha$ emission may be blue-shifted relative to the systemic redshift of the QSO.
We therefore consider our estimate with caution, but note that previous work indicating a systemic redshift
of $z_{\rm em}=2.7783\pm 0.007$ \citep{Ellison2010} would lead to an even larger velocity ($\sim 3500$~\kmps).

The observed metal-line profiles are complex, exhibiting at least $\sim 20$ individual velocity components
with two main clumps at $z_{\rm abs}\approx 2.8114$ and $2.8138$, respectively. Molecular hydrogen is
detected within two components of the bluest clump only, at $z_{\rm abs}=2.810995(2)$ and $2.811124(2)$, with
total column densities of $\log N($H$_2)=18.10\pm 0.02$ and $17.82\pm 0.02$,
respectively \citep{Klimenko2015}. We adopt these values in the following analysis as they were derived from
the same spectrum. Deuterated molecular hydrogen (HD) is detected in lines originating from
the $J=0$ rotational level corresponding to the H$_2$ component at $z_{\rm abs}=2.811124$. The total HD column
density is $\log N=13.33\pm 0.02$ \citep{Klimenko2015}. Throughout the paper, we use $z=2.811124$ to define the zero of the velocity scale.

\begin{table}
\caption{Fit results for \ion{Si}{ii}$^\star$ lines, originating from the first excited level of singly-ionized
silicon.}
\centering
\begin{center}
\begin{tabular}{ccc}
\hline
Component & $A$ & $B$ \\
\hline
$z$                            & 2.810973 & $2.811129(3)$           \\
$\Delta v^\dagger$ [km~s$^{-1}$]       & -$11.9$              & $0.4^{+0.2}_{-0.2}$                    \\
$b$        [km~s$^{-1}$]       & 3.8$^\ddagger$      & $3.5^{+0.5}_{-0.6}$     \\
\hline
$\log N($\ion{Si}{ii}$^\star)$ & $<11$            & $11.37^{+0.03}_{-0.03}$ \\
\hline
\end{tabular}
\end{center}
\label{table:SiII_exc_results}
\begin{tablenotes}
\item $^\dagger$ Velocity offset relative to $z=2.811124$. 
\item $^\ddagger$ In this component, the redshift and Doppler parameter are taken from the
fit of the C\,{\sc i} lines, and were fixed to these values when fitting
\ion{Si}{ii}$^\star$ to determine an upper limit on its column density.
\end{tablenotes}
\end{table}

\begin{figure*}
\centering
\includegraphics[trim={0.0cm 0.0cm 0.0cm 0.0cm},clip,width=\textwidth]{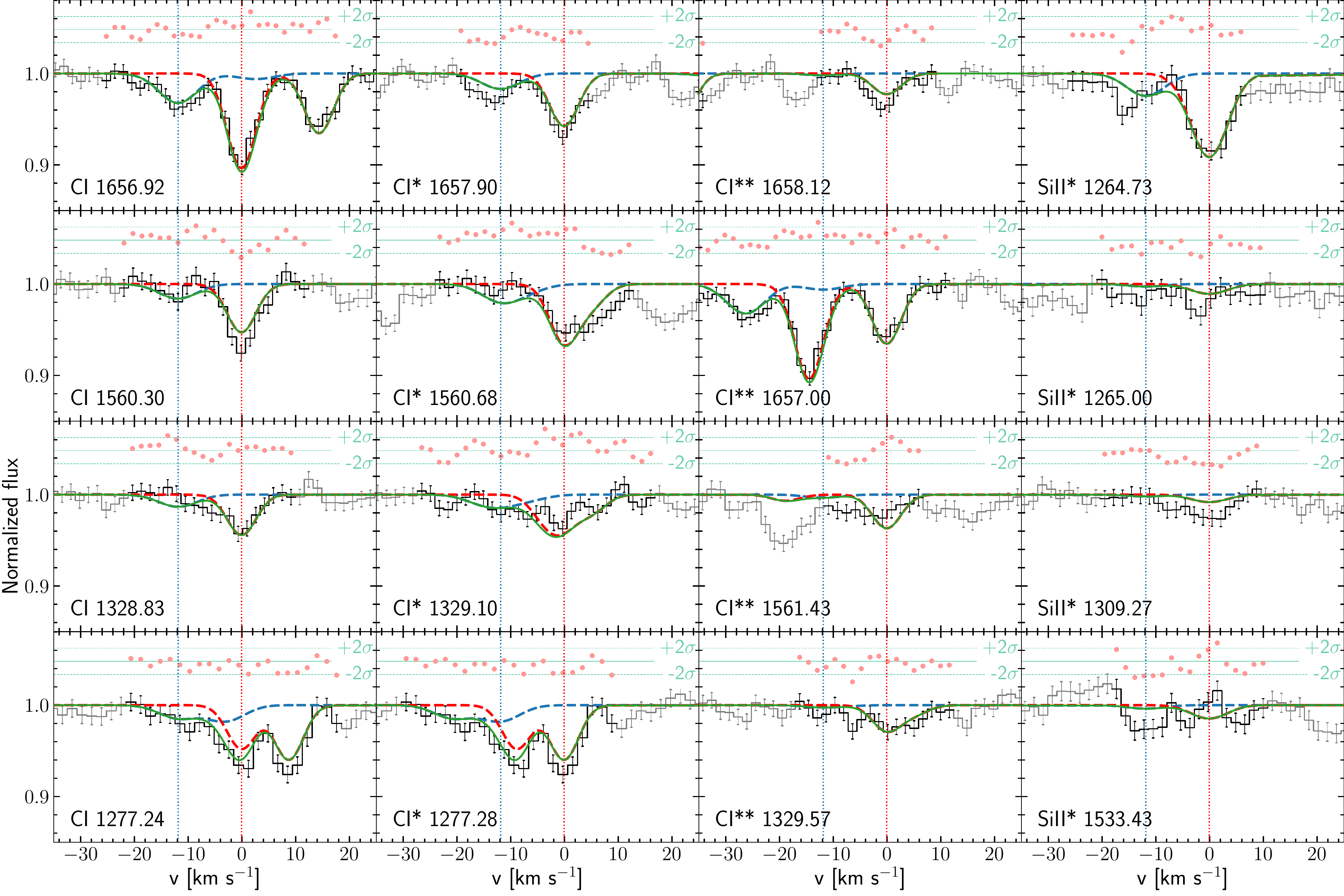}
\caption{Voigt-profile fits to \ion{C}{i} and \ion{Si}{ii} lines at $z_{\rm abs}=2.811$
towards \qso. The observed spectrum is shown in grey except for those pixels used to
constrain the model which are displayed in black. The blue and red dashed curves are the fits
of components $A$ and $B$, respectively, and their joint contribution is shown by green solid line.
Fit residuals are displayed in each panel above the spectra. The zero-velocity of
the x-axis corresponds to a redshift of $z=2.811124$.
\label{fig:CI_profiles}}
\end{figure*}

\subsection{\ion{C}{i} fine-structure levels}

Absorption lines corresponding to transitions from the three fine-structure energy levels of the \ion{C}{i}
ground state are detected in two components that coincide in velocity with the detected H$_2$ components. In the following, we will refer to the component near the zero-velocity redshift as $B$ (or red) and the component bluewards of it as $A$ (or blue). We
use the transitions around 1656, 1560, 1328, and 1277~\AA, to measure the column densities in all
three \ion{C}{i} fine-structure levels using a two-component model with tied Doppler parameters. This
implicitly assumes that the medium associated with each component is
homogeneous.
The fitted \ion{C}{i}-line profiles are shown in Fig.~\ref{fig:CI_profiles} and the fit parameters are given
in Table~\ref{table:CI_results}. The Doppler parameter derived for the blue component is consistent with
that obtained from H$_2$ \citep[see][]{Klimenko2015}. However, we find higher Doppler parameter for the
red \CI\ component than in lower H$_2$ rotational levels. Notwithstanding, we note that \ion{C}{i} lines
are weak so the derived column densities (and especially their ratios) are not very sensitive to the exact
value of the Doppler parameter. The column densities derived in both components indicate a relatively high
\ion{C}{i} excitation compared to what is usually seen in intervening high-$z$ DLAs: $J=1$ is in the present
case more populated than $J=0$, when the opposite is normally seen \citep[see, e.g.,][]{Jorgenson2010, Balashev2011, Noterdaeme2017}. The observed column-density ratios is used
in Sect.~\ref{sect:PhysCond} to estimate the hydrogen number density of the absorbing medium.

\subsection{\ion{Si}{ii} fine-structure levels}
\label{sect:SiII}
We detect ionized silicon in its first excited state (\ion{Si}{ii}$^\star$) from its strong absorption line at rest-frame wavelength 1264~\AA.  This absorption is confidently detected in the component coinciding with the red component of the H$_2$- and \ion{C}{i}-line profiles, while some absorption can only be hinted at the position of the strongest metal lines. The fitted \ion{Si}{ii}$^\star$ line is shown in the right-most panels of Fig.~\ref{fig:CI_profiles} and the corresponding fit parameters are given in Table~\ref{table:SiII_exc_results}. For component $B$, we place an upper limit on the column density assuming the redshift and Doppler parameter obtained from \ion{C}{i}.

It is not possible to accurately determine the \ion{Si}{ii} column density associated with
the H$_2$-detected components of the DLA since the only \ion{Si}{ii} absorption line which
is not completely saturated (i.e., \ion{Si}{ii}$\lambda 1808$) is blended with telluric
absorption. Additionally, many velocity components in the intermediate-saturation regime
of the \ion{Si}{ii}$\lambda 1808$ line are blended with each other, which increases
the degeneracy to the fitting solution as derived from a single line. Therefore, we
roughly estimate the \ion{Si}{ii} column density corresponding to the
H$_2$-detected component using the measured \ion{Zn}{ii} column densities. For this, we
fit \ion{Zn}{ii}, \ion{Cr}{ii}, \ion{Ni}{ii}, and \ion{Fe}{ii} lines assuming the same redshift and Doppler parameter for each metal species within a given component. We provide
the fit results in Table~\ref{table:metal} of the Appendix and show the fits
in Figs.~\ref{fig:metal_profiles_2} and \ref{fig:metal_profiles_3}. We
measure \ion{Zn}{ii} column densities of $\log N = 11.99^{+0.05}_{-0.04}$ and
$12.28^{+0.01}_{-0.02}$ in the components closest to the blue and red
H$_2$-detected components, respectively. The total \ion{Zn}{ii} column density is
$\log N = 13.25^{+0.02}_{-0.02}$, yielding an average DLA metallicity (calculated using
solar abundances from \citealt{Asplund2009}) based on \ion{Zn}{ii} of
${\rm [X/H]}=-0.68^{+0.02}_{-0.02}$. This metallicity is $\sim$0.2 dex larger than
that derived by \citet{DeCia2016} due to our more sophisticated decomposition of
this complex line profile. We find that the depletion of Cr, Ni, and Fe relative to
Zn is $-0.47^{+0.01}_{-0.01}$, $-0.57^{+0.01}_{-0.01}$, and $-0.76^{+0.01}_{-0.01}$,
respectively, based on total column densities (which is also similar on a
component-by-component basis). This suggests that the depletion of Si relative to Zn
is $\sim 0.1$ with a dispersion of around $0.15$ \citep{DeCia2016}. Therefore, we
estimate column densities of $\log N(\ion{Si}{ii}) = 14.84^{+0.15}_{-0.15}$
and $15.13^{+0.15}_{-0.15}$, in the blue and red H$_2$-detected components, respectively.

\section{Physical conditions}
\label{sect:PhysCond}

We use the observed excitation of the fine-structure energy levels of the \ion{C}{i} and \ion{Si}{ii}
ground states to constrain the physical conditions in the cold-gas phase probed by the H$_2$/\ion{C}{i}-detected
components of the DLA. For this, we perform standard calculations alike \textsc{POPRATIO} \citep{Silva2001},
the details of which, and information on the data we use, is given in \citet{Balashev2017}. Our calculations
assume a homogeneous medium where \ion{C}{i} and \ion{Si}{ii} fine-structure levels are populated by
collisions with atomic and molecular hydrogen\footnote{We also take into account Helium which, however, only has
a minor contribution to the excitation. On the other hand, we neglect collisions with electrons due to the
low-ionization fraction of the Cold Neutral Medium associated with the H$_2$/\ion{C}{i} components.}, and pumping
by UV photons\footnote{\ion{Si}{ii} and \ion{C}{i} fine-structure levels can be populated by direct radiative
excitation. This kind of excitation requires very strong IR/sub-mm fields. Principally, such photons can
be supplied by AGN. However, we checked that for a typical broad-band AGN spectrum, UV pumping dominates over
direct excitation by several orders of magnitude both for \ion{Si}{ii} and \ion{C}{i}.}. The cosmic
microwave background radiation intensity is fixed by the redshift of the DLA. The observed molecular fraction
yields the fraction of \ion{H}{i}/H$_2$ -- two main collisional partners.
We assume that hydrogen is predominantly in atomic form since the H$_2$ column density is too low for H$_2$ to
be completely self-shielded. PDR Meudon code calculations for the derived physical conditions strongly support this
assumption. We are hence left with three parameters which altogether characterize the excitation of \ion{C}{i}
and \ion{Si}{ii}: the hydrogen number density\footnote{In our case, since we consider a neutral medium with low molecular fraction, hydrogen is mostly in atomic
form and therefore $n_{\rm H} = n_{\rm \ion{H}{ii}} + n_{\rm \ion{H}{i}} + 2 n_{\rm H_2} \approx n_{\rm \ion{H}{i}}$.}, $n_{\rm H}$, the UV radiation-field strength, and the temperature, $T$. In contrast to \ion{C}{i} whose lines are optically thin, the \ion{Si}{ii} lines corresponding to the strongest transitions that participate in UV pumping are optically thick. We take into account the reduction of  UV pumping for \ion{Si}{ii} using the estimated column densities in components $A$ and $B$ (see Sect.~\ref{sect:SiII}). We build a
grid of values for each of these parameters, within reasonable ranges, and compare calculated relative populations
with the observed ones, assuming Gaussian probability distributions for the measurement uncertainties.


To simplify the representation of the results, we perform calculations separately onto two 2D planes of parameter
space, $n_{\rm H}-T$, and $n_{\rm H}-\rm UV$, respectively. This is justified since the kinetic temperature is well constrained
from $T_{01}$, which is the excitation temperature estimated from the relative populations of the H$_2$ $J=1/J=0$
rotational levels, which was previously found to be $\rm T_{01}=141\pm6$\,K and $167\pm13$\,K in components
$A$ and $B$, respectively \citep{Klimenko2015}. Additionally, we found that pumping by UV photons has a minor
effect on the populations of \ion{C}{i} fine-structure levels.

\begin{figure*}
\centering
\setlength{\tabcolsep}{2pt}
\begin{tabular}{cc}
\includegraphics[trim={0.0cm 0.0cm 0.0cm 0.0cm},clip,width=0.48\textwidth]{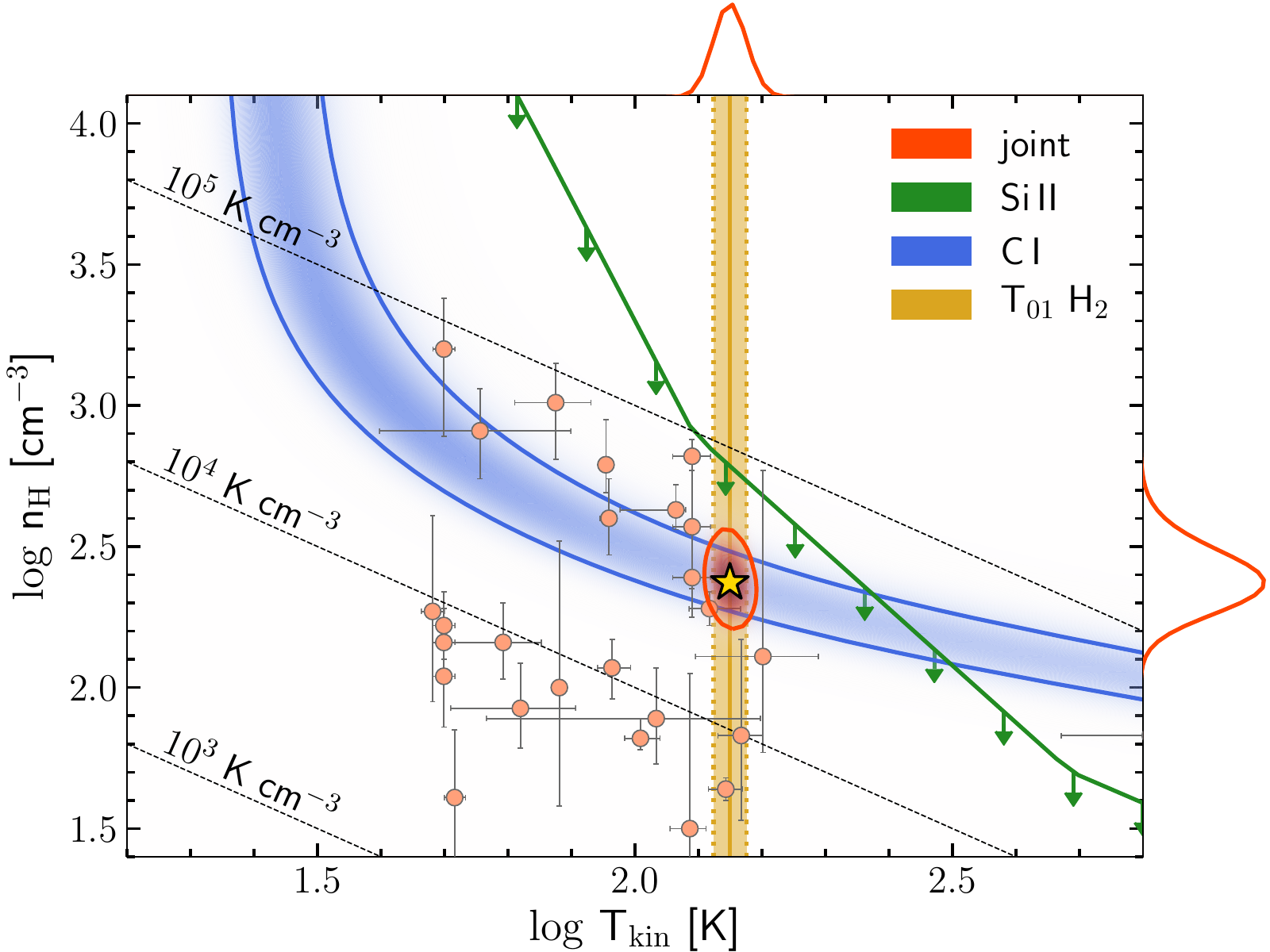}
& \includegraphics[trim={0.0cm 0.0cm 0.0cm 0.0cm},clip,width=0.48\textwidth]{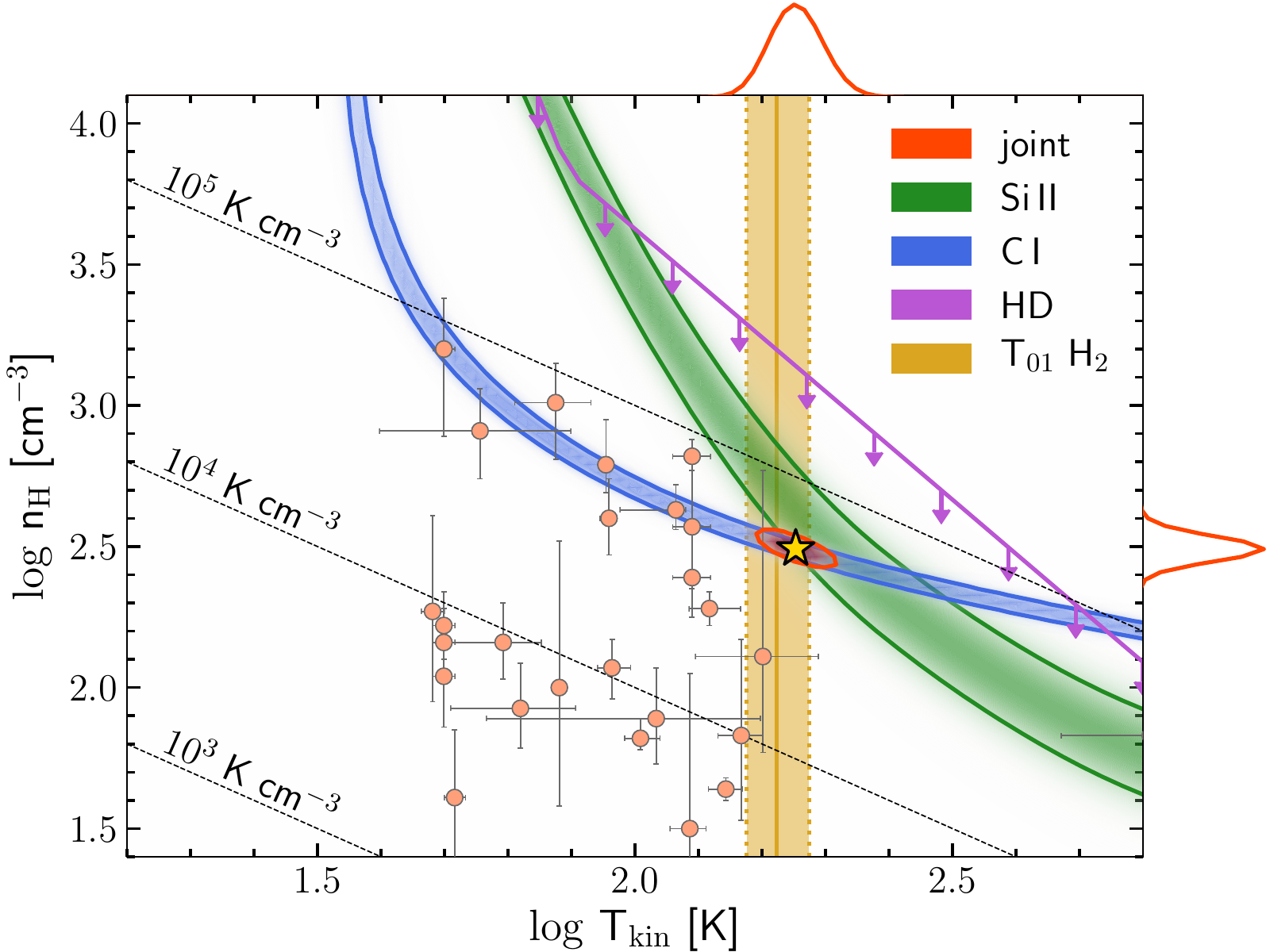}\\
\end{tabular}
\setlength{\tabcolsep}{6pt}
\caption{Estimates of hydrogen number density and temperature in component $A$ (left panel) and $B$
(right panel) of the DLA, associated with the H$_2$/\CI-bearing gas. The green, blue, violet, and yellow
colors are used to show the constraints from \ion{Si}{ii}, \ion{C}{i}, HD, and $T_{01}$ (i.e., the
H$_2$ ortho-to-para ratio), respectively. Color gradients depict the derived probability-density
functions, with contour lines encompassing 68.3\% of them. The red contour in each panel is the joint constraint, with a
yellow star marking the location of the most likely value. The red curves on the top and right-hand side axes show
the marginalized distributions arising from the joint fit of $T_{\rm kin}$ and $n_{\rm H}$, respectively. Light
red circles are estimates of $n_{\rm H}$ and $T_{01}$ in generic H$_2$/\ion{C}{i}-bearing DLAs
at high redshift \citep[][]{Balashev2019}.
}
\label{fig:phys_cond_A}
\label{fig:phys_cond_B}
\end{figure*}

In Fig.~\ref{fig:phys_cond_A}, we show the constraints on number density and temperature inferred from \ion{C}{i},
\ion{Si}{ii}, HD, and $T_{01}$, whilst neglecting UV pumping, in components $A$ and $B$ individually. In
component $B$ at $T\sim T_{01}$ the estimate based on \ion{Si}{ii} provides higher number density than \ion{C}{i}
does. As shown below, this is due to the fact that for \ion{Si}{ii} UV pumping cannot be neglected.
Joint estimates from both species assuming $T_{\rm kin}=T_{01}$ are shown as red contours
in Fig.~\ref{fig:phys_cond_A}, yielding number densities of $240^{+70}_{-60}$ and $310^{+40}_{-30}$\,cm$^{-3}$
in the blue and red components, respectively. HD molecules in the $J=0$ rotational level were previously
detected in the red component with column density $\log N=13.33\pm0.02$ \citep{Klimenko2015}. Here, we derive in
this particular component an upper limit on the column density of HD in the $J=1$ rotational level of
$\log N<12.90$. In typical conditions of the cold ISM, the HD $J=1$ rotational level is mainly populated
by collisions \citep{Balashev2010}. The constraint, in the temperature-density plane, resulting from the
upper limit on the relative populations of the HD $J=1/J=0$ rotational levels, using the same code as above,
is shown (by a purple line) in Fig.~\ref{fig:phys_cond_B} (right panel). It is consistent with previous
estimates from both \ion{C}{i} and \ion{Si}{ii}.

\begin{figure*}
\centering
\setlength{\tabcolsep}{2pt}
\begin{tabular}{cc}
\includegraphics[trim={0.0cm 0.0cm 0.0cm 0.0cm},clip,width=0.48\textwidth]{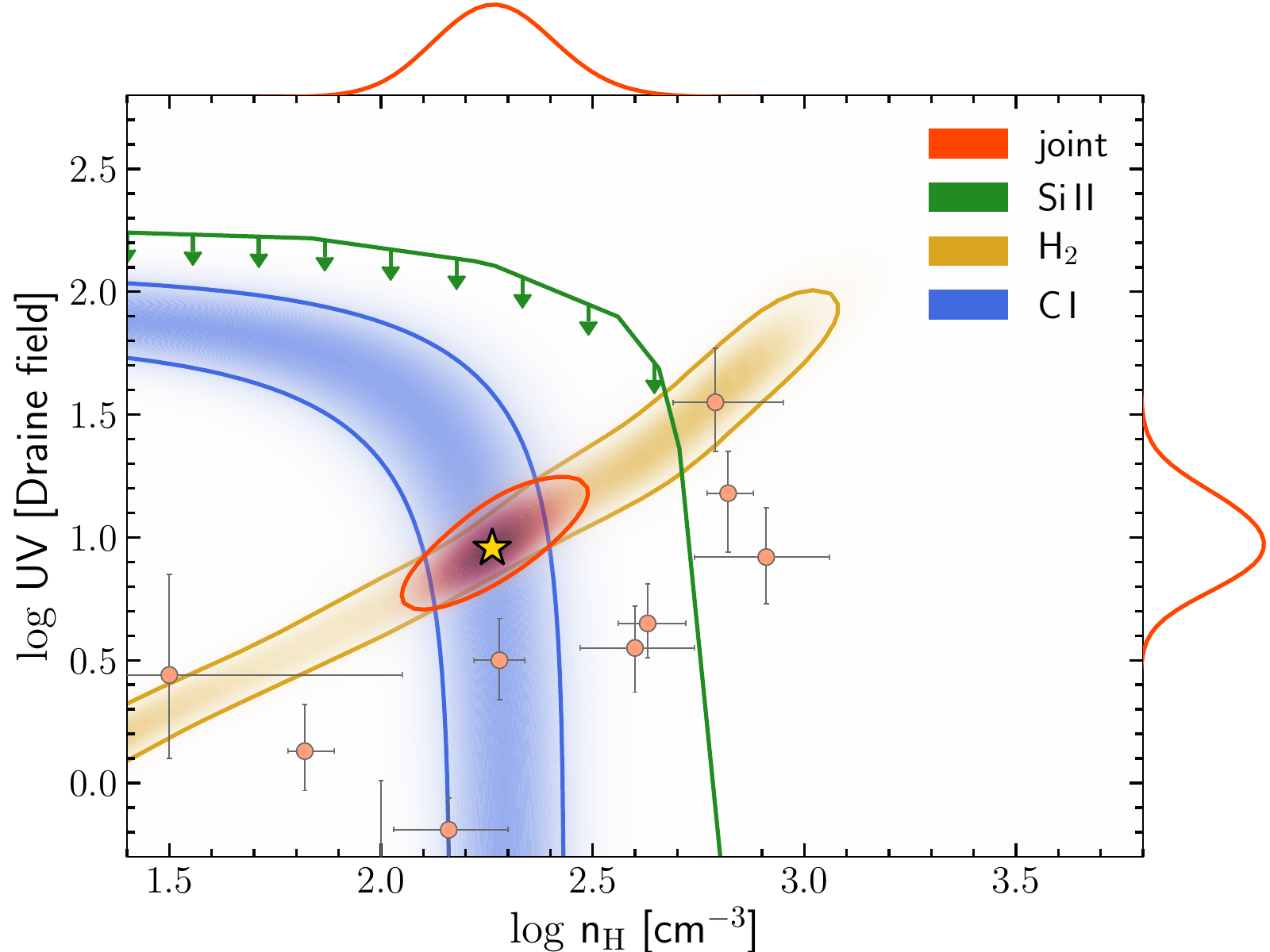}
& \includegraphics[trim={0.0cm 0.0cm 0.0cm 0.0cm},clip,width=0.48\textwidth]{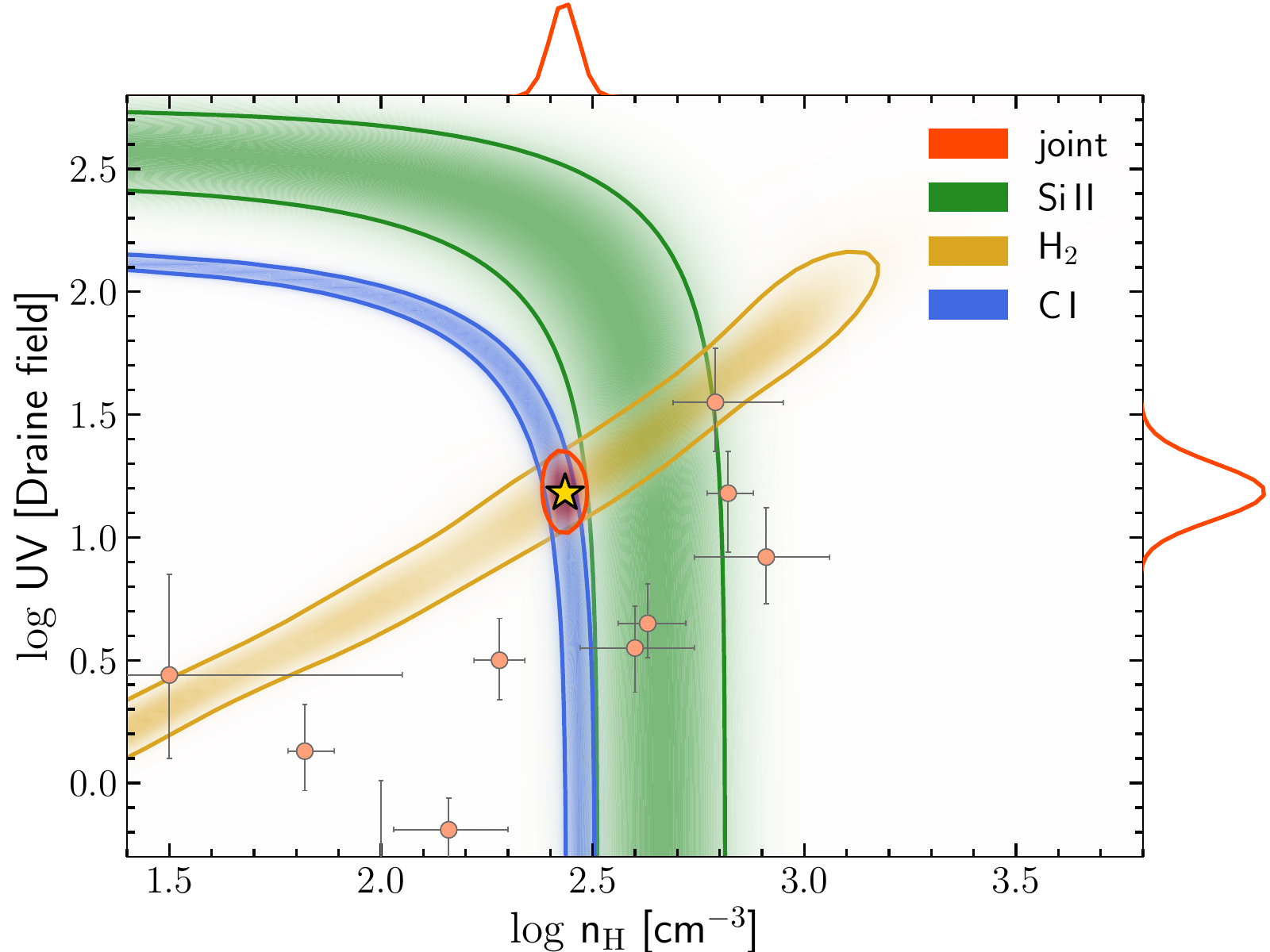}\\
\end{tabular}
\setlength{\tabcolsep}{6pt}
\caption{Estimates of hydrogen number density and UV flux in component $A$ (left panel) and $B$ (right panel)
of the DLA, associated with the H$_2$/\CI-bearing gas, using constraints
from \ion{Si}{ii}, \ion{C}{i}, and H$_2$. Drawing conventions are the same as
in Fig.~\ref{fig:phys_cond_A}. The constraints arising from the excitation of H$_2$ rotational levels
are displayed in yellow.
}
\label{fig:phys_cond_uv_A}
\label{fig:phys_cond_uv_B}
\end{figure*}

We present in Fig.~\ref{fig:phys_cond_uv_A} the constraints on number density and UV flux in components $A$ and
$B$ based on the excitation of \ion{C}{i} and \ion{Si}{ii} assuming $T_{\rm kin} = T_{01}$ in our calculations.
We also use the measured excitation of high H$_2$ rotational levels \citep[taken from][]{Klimenko2015} to further
restrain the solutions \citep[see][]{Balashev2019}. Indeed, H$_2$ rotational levels are populated by
collisions and radiative pumping in resonant lines in a UV wavelength range similar to \ion{C}{i} and
\ion{Si}{ii}. However, resonant lines quickly saturate when external UV radiation penetrates the cloud and
therefore the excitation of H$_2$ rotational levels strongly depends upon the radiative transfer of resonant UV
lines in the H$_2$ Lyman and Werner bands, in which case one cannot assume homogeneity. We hence use in this case
the PDR Meudon code \citep{LePetit2006} that calculates the full radiative transfer in H$_2$ lines to compare
modeled excitation of H$_2$ rotational levels with the observations. We use pseudo-spherical models of
constant density, and metallicity and dust content of one tenth of solar. As for \CI\ and \SiII, we end up
with two external parameters: the hydrogen number density, $n_{\rm H}$, and the strength of the UV radiation
field. We run a grid of models with 10 points evenly spaced logarithmically for both the number density and the UV
field, varied within the following ranges: $\log n_{\rm H} = 1..4$ and $\log \rm UV = -0.5..3$ (where UV
is expressed in units of the Draine's field), respectively. When comparing model results with the observations,
we take into account the column densities measured in all detected H$_2$ rotational levels up to $J=5$. We also use
the column densities of H$_2$ rotational levels calculated at a depth in the cloud corresponding to half of the
total observed H$_2$ column density. This essentially simulates the result of a cloud illuminated on both sides,
with a total H$_2$ column density equal to the measured one. We include a factor of 2 (0.3~dex) uncertainty on
the observed H$_2$ column densities to take into account geometric effects since the H$_2$ cloud is most likely
not exactly spherical. We smooth the results of our calculations by interpolating the H$_2$ excitation diagram on
a denser grid in the $\log n_{\rm H}$-$\log \rm UV$ parameter space, and compare this to the
observed H$_2$ population in Fig.~\ref{fig:phys_cond_uv_A}. We find that the H$_2$, \ion{C}{i}, and \ion{Si}{ii}
2D-probability density functions individually cover a large area of the parameter space but have different shapes
and intersect in a localised region. Joint estimates (shown in the figure by red contours) provide the following
constraints on the number density and UV field strength: $n_{\rm H} = 190^{+70}_{-50}$\,cm$^{-3}$ (resp.
$260^{+30}_{-20}$\,cm$^{-3}$) and $\rm UV = 10^{+5}_{-3}$ (resp. $14^{+3}_{-3}$) in component $A$ (resp. component
$B$). We note that at given joint constraints \ion{C}{i} is predominantly populated by collision, while
\ion{Si}{ii} is excited by UV pumping and collisions.

\section{Discussion}
\label{sect:discussion}

In Figs.~\ref{fig:phys_cond_A} and \ref{fig:phys_cond_uv_A}, we give estimates of number density, temperature,
and UV flux in a sample of H$_2$/\ion{C}{i}-bearing DLAs, which we also determined using our code from
the excitation of \ion{C}{i} fine-structure and H$_2$ rotational levels \citep[i.e., the compilation of data
from][]{Balashev2019}. One can see that the UV flux in both components of the DLA towards \qso, as well as the
thermal pressure (shown by inclined dashed lines in the figure), are higher than the average of other
H$_2$-bearing DLAs. \citet{Balashev2017, Balashev2019} also reported tentative evidence of an increase of
thermal pressure with increasing hydrogen column density. This suggests that high \HI\ column-density systems
have smaller impact parameters than typical DLAs and hence probe the inner regions of galaxies where the thermal
pressure may be enhanced by high UV flux.\footnote{This is in line with the findings of \citet{Noterdaeme2014} and
\citet{Ranjan2019} based on, respectively, statistical and direct detection of star-formation signatures
(which correlate with UV flux) at no more than a few projected kpc from extremely strong
DLAs ($\log N($H\,{\sc i}$)\sim 22$).} However, the DLA studied here has significantly higher thermal
pressure than suggested by this trend for $\log N(\mbox{\rm \HI})\sim 21.35$. We suggest that the relatively
high pressures and high UV fluxes observed in this DLA are instead due to its proximity to the quasar.



\subsection{Origin of the DLA gas}

From a constrained UV flux of about 10-15 times the Draine's field and the known quasar
luminosity, we can estimate the distance between the H$_2$-detected components of the
DLA and the AGN assuming the quasar dominates the incident UV flux. Corrections due to
dust reddening are negligible since the DLA has a low metallicity and we do not see
reddening in the quasar spectrum. Under these circumstances, we infer a distance
of $\sim 150-200$\,kpc.
This distance 
suggests that the DLA is located with a group of
galaxies rather than being produced by the quasar-host galaxy itself. However, such a
distance does not exclude the possibility that the DLA gas is compressed by the
mechanical output of the AGN. Indeed, \qso\ is a blazar, i.e., the line-of-sight where
the absorption system is seen coincides with the direction of the jet, which can have an
extension comparable to the distance we derive \citep[see, e.g.,][]{Blandford2019}.
However, the redshift of the DLA is higher than that of the quasar which implies that
the DLA is moving towards the quasar and cannot be outflowing material.

As mentioned in Sect.~\ref{sect:history}, emission-line searches have revealed
three Ly$\alpha$ emitters around \qso\ that span $\sim100$\,kpc in the transverse
direction. One object is located at an impact parameter of $\sim 8$\,kpc from the
quasar \citep{Warren1996} while the object at the largest impact parameter was
resolved into two sources from HST imaging \citep{Moller1998}. The Ly$\alpha$
emission-line profiles are broad and it is difficult to firmly establish the association
of any of those with the DLA. However, the most reasonable assumption is that the DLA
absorption is associated with the object at the lowest impact parameter, which is corroborated by many lines of evidence \citep[e.g.,][and references therein]{Krogager2017}.
Since we have estimated the distance between the DLA and the QSO to be $\lesssim 200$\,kpc,
which is similar to the transverse separation between the detected Ly$\alpha$
emission sources, these sources probably form a group of which the QSO-host galaxy is a
major member.

Additionally, low-ionization metal lines associated with the DLA exhibit complex kinematics
with two prominent velocity clumps separated by $\sim 200$\,km~s$^{-1}$. This
may suggest that the overall absorption originates from two compact
galaxies and/or includes tidally-stripped gas due to galaxy
interaction \citep[see, e.g.,][]{Kacprzak2010}. We cannot disentangle between
these possibilities as we measured the physical conditions and constrained the distance
of the gas from the H$_2$/\CI\ components of the bluer clump only. However, this is in line
with previous findings \citep[e.g.,][]{Ledoux2006,Christensen2014} where the correlation
between gas velocity dispersion and metallicity in DLAs has been interpreted as a result
of increasingly more massive and complex environments. This is also supported by
recent ALMA observations of J\,1201$+$2117 from which extreme DLA kinematics has
been proposed as the signpost of major mergers in normal galaxies at high redshift
\citep{Prochaska2019}.

\subsection{Physical extent of $\rm H_2$-bearing clouds and partial covering}

Using derived number densities, we can estimate the physical extent of the clouds along the line-of-sight.
From PDR modeling, we found that \ion{H}{i} is the dominant species within the
H$_2$/\ion{C}{i}-detected components. However, one cannot directly measure how much \ion{H}{i} is associated with
each H$_2$ component since individual \HI\ lines are unresolved. This can however be estimated
indirectly. Firstly, we derived a column density $\log N\sim 20$ using the Meudon PDR code using the measured
H$_2$ column densities as input, a UV flux $\sim 15$ times stronger than Draine's field, and a number density
$n_{\rm H}\sim 300$\,cm$^{-3}$ (our derived best-fit parameters). Secondly, the \ion{Zn}{ii} column density
associated with the H$_2$ components is about 1.2~order of magnitude smaller than the overall zinc column
density. Assuming the metallicity is the same in each component of the DLA, we estimate the \HI\ column density
associated with the H$_2$ components to be $\log N\sim 20.1$. \citet{Srianand2012} estimated a similar \HI\
column density in the cold phase of the ISM (i.e., $\log N\sim 20$) based on an upper limit on $T_s / f$ from
non-detection of \HI\ 21-cm absorption.

The most conservative estimate of the total column density of particles (atomic and molecular hydrogen altogether)
associated with the two H$_2$ components of the DLA ranges between $\log N\sim 21.30$ and $\sim 18.0$. Using
these values, the upper and lower bounds on the physical extent of the clouds are $<0.1$\,pc and
$>100$\,a.u., respectively. Assuming a value of $\log N\sim 20$ (see above), we infer a typical size for the
clouds of $\sim 10^4$\,a.u.

The small physical extent of the cloud and/or the jet orientation provide natural explanations for
the $\sim2\%$ line-flux residual seen in the core of the saturated H$_2$ lines \citep{Klimenko2015}. Indeed, this
suggests the continuum source of the quasar is not fully covered by the H$_2$-bearing clouds. Partial covering can
be explained either by (i) comparable sizes of the accretion disc and the H$_2$ absorbing medium; (ii) jet-induced
scintillation in a clumpy medium; (iii) emission from the quasar- and/or the DLA-absorbing galaxy \citep{Cai2014}.

Studying the possible variation with time of the residual flux may provide clues to discriminate between
these scenarios since each of them occur on different timescales.
In the first scenario (geometric effects), flux variations are expected due to the relative projected motion
of the absorber, quasar-emitting region, and observer \citep{Boisse2015}. The velocity of this motion can be
estimated as $\sim 2000$\,km~s$^{-1}$ in the case of the DLA towards \qso\ based on the velocity
difference between QSO and DLA projected on the line-of-sight. This corresponds to a transverse shift of
$\sim 400$\,a.u. in a year. Over 25 years (the available time span of available observations), this is
comparable to the size of the absorber assuming the transverse size of the H$_2$ system is the same as
that derived longitudinally.
Under the second scenario (jet scintillation), the timescale and magnitude of the expected variation of
flux residuals correspond to the quasar variability itself since observed residuals are scattered emission.
However, since the exact position of the scintillated cloud is unknown, the timelag between quasar continuum
and scintillated emission is uncertain making quantitative conclusions from this observation complicated.
Under the third scenario, variability of residual flux is not expected over timescales smaller than a few decades.


\section{Conclusions}
\label{sect:conclusion}

We presented the detection of excited fine-structure energy levels of singly ionized
silicon and neutral carbon associated with the $z_{\rm abs}=2.811$ PDLA towards \qso. We
measured total \ion{Si}{ii}$^\star$ and \ion{C}{i} column densities of $11.37\pm 0.03$ and
$12.71\pm 0.02$, respectively. The metallicity of the DLA
is $\log Z/Z_{\rm \odot}=-0.68\pm 0.02$ based on [\ion{Zn}{ii}/\ion{H}{i}]. From
the relative populations of the fine-structure levels of \ion{Si}{ii} and \ion{C}{i},
as well as the populations of H$_2$ rotational levels, we were able to constrain the
hydrogen number density and UV flux in the cold neutral phase of this DLA. We found number
densities $n_{\rm H}=190^{+70}_{-50}$\,cm$^{-3}$ and $260^{+30}_{-20}$\,cm$^{-3}$ in the
blue and red velocity components, respectively, where both \ion{C}{i} and H$_2$ are
detected. The strengths of the UV field in Draine's unit were derived to
be $I_{\rm UV} = 10^{+5}_{-3}$ and $14^{+3}_{-3}$ in each of the two
components, respectively. The high thermal pressure in comparison with intervening DLAs and
the enhanced UV flux measured in this particular DLA are most likely due to the
proximity of the quasar. Assuming the UV flux is dominated by the quasar, we estimated
the distance between the quasar and the absorbing gas to be $\sim 150-200$\,kpc, and its
physical extent $\sim 10^{4}$\,a.u. This favours a scenario where the absorption occurs in a
companion galaxy located in the group where the quasar-host galaxy resides. This is in line
with studies in emission that revealed the presence of several galaxies around the quasar.

\section{Data availability}
The paper is based on the data that is available on the archive of European Southern Observatory archive under ESO programs 66.A-0594, 68.A-0600, 68.A-0106, and 082.A-0087. 

\section*{Acknowledgements}

We thank the referee for useful comments and suggestions.
This research was partially supported by RFBR grant 18-52-15021. SB is supported by the Foundation for the Advancement of Theoretical Physics and Mathematics ``BASIS'' and
is grateful to the European Southern Observatory in Chile for hospitality as
visiting scientist. We acknowledge support from the French {\sl Agence Nationale de la
Recherche} under ANR grant 17-CE31-0011-01,  project ``HIH2'' (PI: Noterdaeme).




\bibliographystyle{mnras}
\bibliography{references}



\appendix

\section{Metal-line fits}
In Table~\ref{table:metal}, we present the complete results from fitting metal lines in the
DLA at $z_{\rm abs}=2.811$ towards \qso\ using multi-component Voigt profiles and
sampling through the MCMC technique. The observed and synthetic profiles of \ion{Zn}{ii},
\ion{Cr}{ii}, \ion{Ni}{ii}, and \ion{Fe}{ii} are shown in
Figs.~\ref{fig:metal_profiles_2} and \ref{fig:metal_profiles_3}.

\begin{figure*}
\centering
\includegraphics[trim={0.0cm 0.0cm 0.0cm 0.0cm},clip,width=0.98\textwidth]{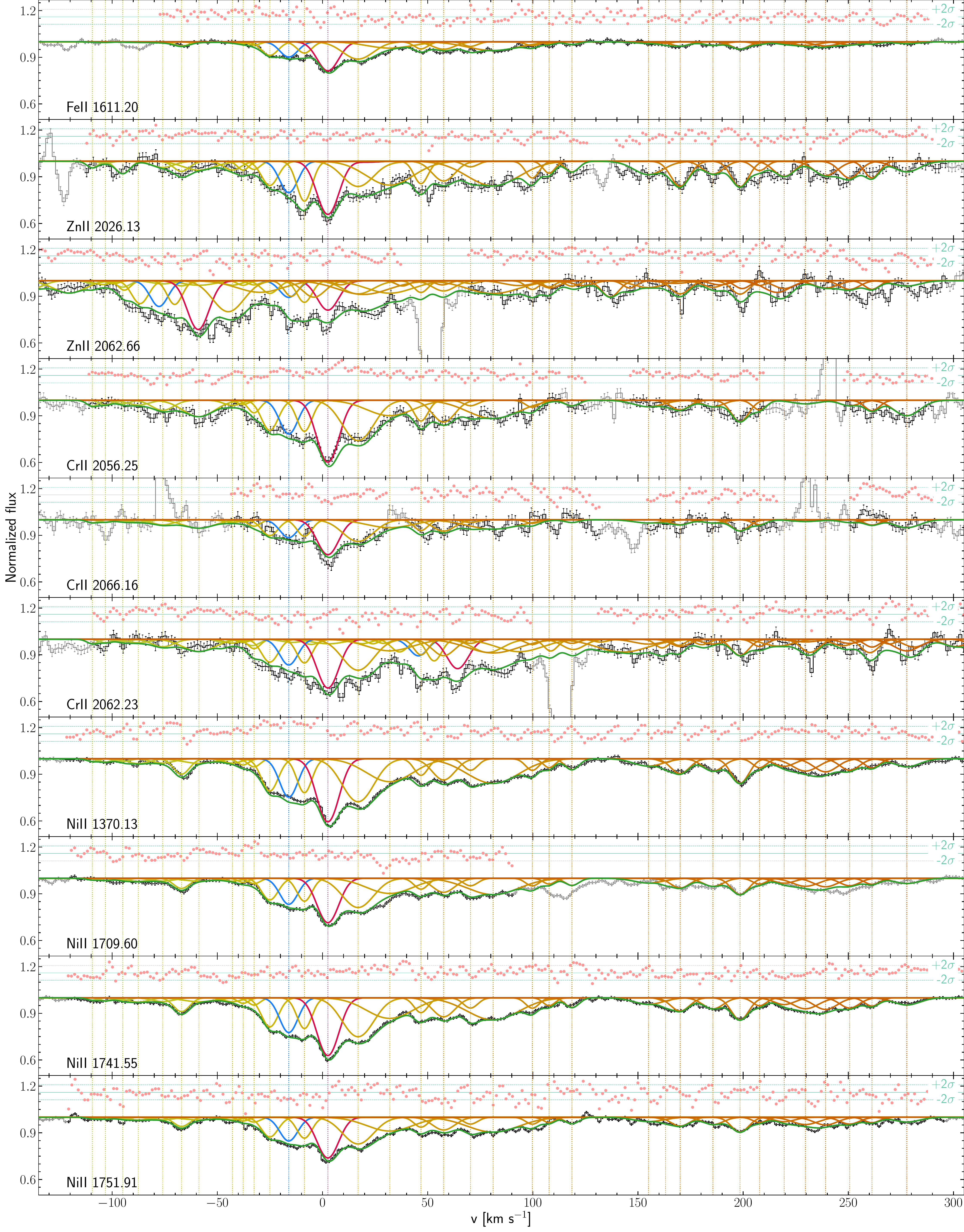}
\caption{Voigt-profile fits to \ion{Zn}{ii}, \ion{Cr}{ii}, \ion{Ni}{ii},
and \ion{Fe}{ii} lines at $z_{\rm abs}=2.811$ towards \qso.
The observed spectrum is shown in grey except for those pixels used to constrain the
model which are instead displayed in black. The total fit profile is represented by
the green curve. Components $A$ and $B$ are displayed in blue and red, respectively,
and the other sub-components are shown in yellow and orange. Fit residuals are displayed in
each panel above the spectra. The zero-velocity of the x-axis corresponds to a redshift
of $z=2.811124$.
\label{fig:metal_profiles_2}}
\end{figure*}

\begin{figure*}
\centering
\includegraphics[trim={0.0cm 0.0cm 0.0cm 0.0cm},clip,width=0.98\textwidth]{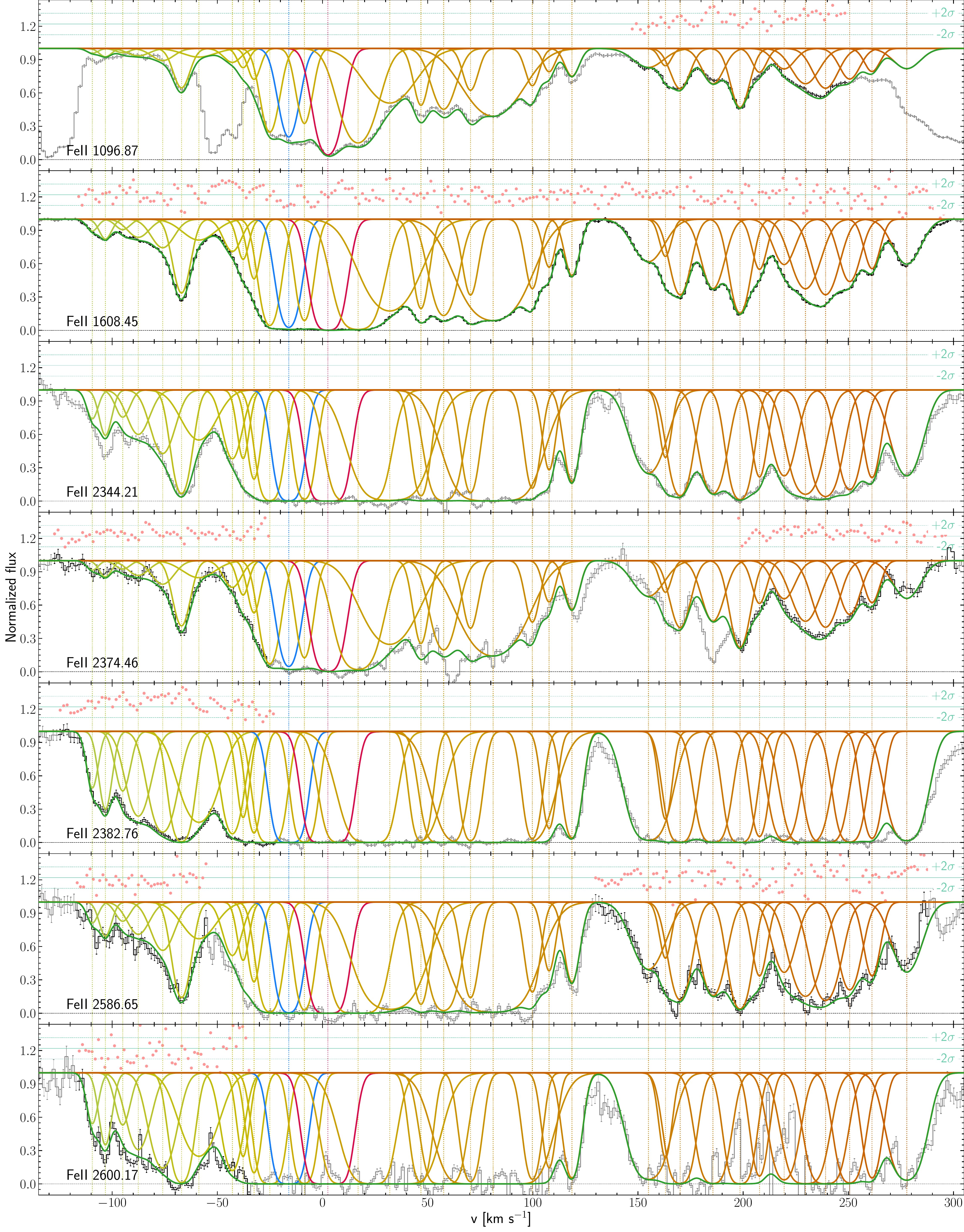}
\caption{Continuation of Fig.~\ref{fig:metal_profiles_2}.
\label{fig:metal_profiles_3}}
\end{figure*}

\begin{table*}
\centering
\caption{Result of Voigt-profile fitting to metal lines in the DLA at
$z_{\rm abs}=2.811$ towards \qso.\label{table:metal}}
\begin{tabular}{cccccccc}
\hline
Comp. & $z$ & $\Delta v^\dagger$ [km~s$^{-1}$] & $b$ [km~s$^{-1}$] &
$\log N($\ion{Zn}{ii}$)$ &
$\log N($\ion{Cr}{ii}$)$ &
$\log N($\ion{Fe}{ii}$)$ &
$\log N($\ion{Ni}{ii}$)$ \\
\hline
1 & $2.8097329(^{+22}_{-22})$ & -109.5 & $1.8^{+0.3}_{-0.2}$ & $11.03^{+0.10}_{-0.21}$ & $11.67^{+0.15}_{-0.21}$ & $12.49^{+0.03}_{-0.02}$ & $11.72^{+0.03}_{-0.08}$ \\
2 & $2.8098120(^{+24}_{-17})$ & -103.3 & $2.3^{+0.2}_{-0.2}$ & $<10.8$ & $11.59^{+0.14}_{-0.22}$ & $12.77^{+0.02}_{-0.02}$ & $11.84^{+0.04}_{-0.06}$ \\
3 & $2.809917(^{+3}_{-5})$ & -95.0 & $3.4^{+0.6}_{-0.5}$ & $11.42^{+0.04}_{-0.10}$ & $11.52^{+0.14}_{-0.32}$ & $12.53^{+0.07}_{-0.09}$ & $11.83^{+0.08}_{-0.10}$ \\
4 & $2.810011(^{+6}_{-4})$ & -87.6 & $6.2^{+0.6}_{-0.6}$ & $<10.5$ & $11.84^{+0.13}_{-0.23}$ & $12.95^{+0.05}_{-0.04}$ & $12.24^{+0.03}_{-0.06}$ \\
5 & $2.810158(^{+3}_{-4})$ & -76.0 & $6.5^{+0.2}_{-0.2}$ & $11.29^{+0.04}_{-0.06}$ & $12.19^{+0.06}_{-0.07}$ & $13.19^{+0.01}_{-0.01}$ & $12.21^{+0.02}_{-0.05}$ \\
6 & $2.8102719(^{+7}_{-5})$ & -67.1 & $4.1^{+0.1}_{-0.1}$ & $11.37^{+0.08}_{-0.10}$ & $<10.2$ & $13.64^{+0.01}_{-0.00}$ & $12.64^{+0.01}_{-0.01}$ \\
7 & $2.810377(^{+3}_{-6})$ & -58.8 & $10.0^{+0.4}_{-0.6}$ & $11.61^{+0.09}_{-0.04}$ & $12.57^{+0.04}_{-0.02}$ & $13.20^{+0.03}_{-0.02}$ & $12.39^{+0.03}_{-0.03}$ \\
8 & $2.810579(^{+6}_{-5})$ & -42.9 & $4.8^{+0.6}_{-0.2}$ & $11.34^{+0.10}_{-0.10}$ & $11.79^{+0.13}_{-0.19}$ & $13.16^{+0.03}_{-0.06}$ & $12.12^{+0.06}_{-0.06}$ \\
9 & $2.8106443(^{+43}_{-30})$ & -37.8 & $1.8^{+0.2}_{-0.3}$ & $11.17^{+0.16}_{-0.08}$ & $11.51^{+0.25}_{-0.24}$ & $13.15^{+0.05}_{-0.04}$ & $12.09^{+0.05}_{-0.07}$ \\
10 & $2.810710(^{+3}_{-4})$ & -32.7 & $2.5^{+0.5}_{-0.2}$ & $11.17^{+0.15}_{-0.17}$ & $12.04^{+0.11}_{-0.09}$ & $13.42^{+0.06}_{-0.06}$ & $12.25^{+0.07}_{-0.09}$ \\
11 & $2.8108055(^{+49}_{-18})$ & -25.1 & $4.4^{+0.4}_{-0.3}$ & $11.79^{+0.06}_{-0.05}$ & $12.57^{+0.05}_{-0.04}$ & $14.18^{+0.03}_{-0.04}$ & $13.01^{+0.02}_{-0.05}$ \\
12 & $2.810920(^{+3}_{-3})$ & -16.1 & $4.9^{+0.5}_{-0.4}$ & $11.95^{+0.07}_{-0.04}$ & $12.66^{+0.06}_{-0.04}$ & $14.27^{+0.05}_{-0.04}$ &  $13.13^{+0.04}_{-0.05}$ \\
13 & $2.8110146(^{+27}_{-19})$ & -8.7 & $3.7^{+0.2}_{-0.3}$ & $11.99^{+0.05}_{-0.04}$ & $12.54^{+0.04}_{-0.05}$ & $14.05^{+0.06}_{-0.03}$ & $12.99^{+0.03}_{-0.05}$ \\
14 & $2.8111566(^{+5}_{-10})$ & 2.5 & $6.1^{+0.1}_{-0.2}$ & $12.28^{+0.01}_{-0.02}$ & $13.05^{+0.01}_{-0.02}$ & $14.64^{+0.01}_{-0.01}$ &  $13.46^{+0.01}_{-0.01}$ \\
15 & $2.8113387(^{+25}_{-17})$ & 16.8 & $9.5^{+0.3}_{-0.2}$ & $12.07^{+0.05}_{-0.04}$ & $12.98^{+0.01}_{-0.04}$ & $14.55^{+0.02}_{-0.02}$ & $13.41^{+0.01}_{-0.02}$ \\
16 & $2.811530(^{+8}_{-8})$ & 31.9 & $14.8^{+0.8}_{-0.5}$ & $12.22^{+0.04}_{-0.04}$ & $12.70^{+0.05}_{-0.04}$ & $14.28^{+0.03}_{-0.02}$ &  $13.28^{+0.02}_{-0.03}$ \\
17 & $2.8117193(^{+8}_{-7})$ & 46.8 & $3.3^{+0.2}_{-0.2}$ & $11.68^{+0.04}_{-0.05}$ & $12.33^{+0.04}_{-0.03}$ & $13.82^{+0.02}_{-0.02}$ & $12.62^{+0.03}_{-0.03}$ \\
18 & $2.8118561(^{+16}_{-9})$ & 57.5 & $6.7^{+0.2}_{-0.2}$ & $11.82^{+0.05}_{-0.05}$ & $12.55^{+0.02}_{-0.04}$ & $14.04^{+0.02}_{-0.01}$ & $12.94^{+0.01}_{-0.02}$ \\
19 & $2.8120187(^{+12}_{-15})$ & 70.3 & $3.5^{+0.2}_{-0.2}$ & $11.06^{+0.11}_{-0.23}$ & $11.74^{+0.13}_{-0.16}$ & $13.64^{+0.02}_{-0.02}$ & $12.47^{+0.03}_{-0.03}$ \\
20 & $2.8121555(^{+32}_{-15})$ & 81.1 & $16.6^{+0.2}_{-0.3}$ & $12.28^{+0.03}_{-0.02}$ & $12.82^{+0.02}_{-0.03}$ & $14.47^{+0.01}_{-0.00}$ &  $13.36^{+0.01}_{-0.01}$ \\
21 & $2.8123923(^{+6}_{-11})$ & 99.7 & $3.1^{+0.2}_{-0.1}$ & $11.53^{+0.05}_{-0.05}$ & $11.82^{+0.12}_{-0.07}$ & $13.55^{+0.01}_{-0.01}$ & $12.44^{+0.02}_{-0.02}$ \\
22 & $2.8124936(^{+14}_{-7})$ & 107.7 & $2.3^{+0.2}_{-0.2}$ & $11.33^{+0.06}_{-0.07}$ & $<10.3$ & $13.26^{+0.01}_{-0.02}$ & $12.27^{+0.03}_{-0.01}$ \\
23 & $2.8126321(^{+5}_{-3})$ & 118.6 & $3.3^{+0.1}_{-0.1}$ & $11.34^{+0.07}_{-0.06}$ & $11.73^{+0.12}_{-0.15}$ & $13.40^{+0.00}_{-0.01}$ & $12.43^{+0.01}_{-0.01}$ \\
24 & $2.8130940(^{+34}_{-10})$ & 154.9 & $9.2^{+0.1}_{-0.2}$ & $11.76^{+0.04}_{-0.04}$ & $12.12^{+0.09}_{-0.06}$ & $13.53^{+0.01}_{-0.01}$ &  $12.57^{+0.01}_{-0.02}$ \\
25 & $2.8131968(^{+18}_{-16})$ & 163.0 & $2.0^{+0.3}_{-0.2}$ & $<10.2$ & $<11.2$ & $13.10^{+0.03}_{-0.03}$ & $11.88^{+0.05}_{-0.09}$ \\
26 & $2.8132860(^{+14}_{-9})$ & 170.0 & $5.2^{+0.1}_{-0.2}$ & $11.85^{+0.03}_{-0.03}$ & $12.10^{+0.07}_{-0.05}$ & $13.72^{+0.01}_{-0.01}$ & $12.69^{+0.01}_{-0.01}$ \\
27 & $2.8134840(^{+8}_{-7})$ & 185.6 & $5.9^{+0.1}_{-0.2}$ & $11.58^{+0.04}_{-0.06}$ & $11.96^{+0.09}_{-0.08}$ & $13.69^{+0.01}_{-0.00}$ & $12.65^{+0.01}_{-0.01}$ \\
28 & $2.8136476(^{+4}_{-7})$ & 198.4 & $4.8^{+0.1}_{-0.1}$ & $11.84^{+0.03}_{-0.02}$ & $12.43^{+0.03}_{-0.03}$ & $13.92^{+0.00}_{-0.01}$ & $12.90^{+0.00}_{-0.01}$ \\
29 & $2.8137653(^{+17}_{-11})$ & 207.7 & $3.3^{+0.3}_{-0.1}$ & $11.28^{+0.05}_{-0.12}$ & $11.97^{+0.08}_{-0.06}$ & $13.32^{+0.02}_{-0.01}$ & $12.34^{+0.02}_{-0.03}$ \\
30 & $2.813914(^{+3}_{-4})$ & 219.4 & $6.0^{+0.3}_{-0.4}$ & $11.67^{+0.06}_{-0.03}$ & $<10.5$ & $13.42^{+0.03}_{-0.06}$ & $12.44^{+0.06}_{-0.04}$ \\
31 & $2.814043(^{+7}_{-5})$ & 229.6 & $7.9^{+0.4}_{-0.6}$ & $<10.3$ & $11.80^{+0.26}_{-0.15}$ & $13.79^{+0.04}_{-0.03}$ & $12.75^{+0.04}_{-0.04}$ \\
32 & $2.814165(^{+3}_{-4})$ & 239.1 & $7.0^{+0.5}_{-0.3}$ & $11.69^{+0.04}_{-0.07}$ & $<10.3$ & $13.81^{+0.03}_{-0.04}$ & $12.73^{+0.03}_{-0.05}$ \\
33 & $2.8143106(^{+25}_{-24})$ & 250.6 & $5.6^{+0.3}_{-0.1}$ & $11.69^{+0.05}_{-0.04}$ & $11.87^{+0.10}_{-0.12}$ & $13.59^{+0.03}_{-0.01}$ & $12.60^{+0.02}_{-0.03}$ \\
34 & $2.8144455(^{+12}_{-9})$ & 261.2 & $3.8^{+0.2}_{-0.1}$ & $11.57^{+0.04}_{-0.06}$ & $12.04^{+0.05}_{-0.07}$ & $13.34^{+0.01}_{-0.01}$ & $12.39^{+0.02}_{-0.01}$ \\
35 & $2.8146554(^{+9}_{-4})$ & 277.7 & $8.0^{+0.1}_{-0.1}$ & $11.60^{+0.06}_{-0.05}$ & $12.46^{+0.03}_{-0.03}$ & $13.51^{+0.00}_{-0.00}$ & $12.56^{+0.01}_{-0.01}$ \\
\hline
$\log N_{\rm tot}$ & ... & ... & ... & $13.25^{+0.01}_{-0.01}$ & $13.86^{+0.01}_{-0.01}$ & $15.43^{+0.00}_{-0.00}$ & $14.33^{+0.00}_{-0.00}$ \\
$\rm [X/H]^\ddagger$ & ... & ... & ... & $-0.68^{+0.01}_{-0.01}$ & $-1.15^{+0.01}_{-0.01}$ & $-1.44^{+0.01}_{-0.01}$ & $-1.26^{+0.01}_{-0.01}$ \\
$\rm [X/\ion{Zn}{ii}]^\ddagger$ & ... & ... & ... & $0.00^{+0.01}_{-0.01}$ & $-0.47^{+0.01}_{-0.01}$ & $-0.76^{+0.01}_{-0.01}$ & $-0.57^{+0.01}_{-0.01}$ \\
\hline
\end{tabular}
\begin{tablenotes}
\item $^\dagger$ Velocity offset relative to $z=2.811124$.
\item $^\ddagger$ Using solar abundances from \citet{Asplund2009}.
\end{tablenotes}
\end{table*}



\bsp	
\label{lastpage}
\end{document}